\documentclass[aps,prl,preprint,amsmath,amssymb]{revtex4-1}

\usepackage{graphicx}
\usepackage{dcolumn}
\usepackage{bm}

\graphicspath{{converted_graphics/}}
\begin{document}

\title{Mechanically Compliant Grating Reflectors for Optomechanics}

\author{Utku Kemiktarak, Michael Metcalfe, Mathieu Durand, John Lawall}

\affiliation{National Institute of Standards and Technology, 100 Bureau Drive, Gaithersburg, MD 20899, USA
		\\ Joint Quantum Institute, University of Maryland, College Park, MD 20742, USA.}

\date{\today}

\begin{abstract}
We demonstrate micromechanical reflectors with a reflectivity as large as 99.4\% and
a mechanical quality factor $Q$ as large as $7.8\times10^5$ for optomechanical applications.  The
reflectors are silicon nitride membranes patterned with sub-wavelength grating structures, obviating the need for the many dielectric layers used in conventional mirrors.  We have employed the reflectors in the construction of a Fabry-Perot cavity with a finesse
as high as $F=1200$, and used the optical response to probe the mechanical properties of the
membrane.  By driving the cavity with light detuned to the high-frequency side of a cavity resonance,
we create an optical antidamping force that causes the reflector to self-oscillate at 211~kHz.
\end{abstract}


\maketitle
Optical Fabry-Perot cavities employing mechanically compliant mirrors have been the subject of intense
recent interest for their utility in quantum-limited position measurements~\cite{Tittonen99},
creating nonclassical states of the optical cavity field and mirror~\cite{Bose97,Penrose03}, 
and attempting to optically cool fabricated mesoscopic 
systems into the quantum regime~\cite{ArcizetPRL06,thompson08,Wilson08,Groblacher09}. The optical cavity provides
a means to interrogate the dynamics of a mechanical resonator with extremely high sensitivity,
while simultaneously providing a large circulating optical power whose radiation pressure can
be employed to alter the mechanics.  

The ideal optomechanical element for such work combines low mass, high mechanical quality 
factor $Q$, and near-unity reflectivity.  A small mass confers a large mechanical zero-point
motion, which is directly proportional to the optomechanical coupling~\cite{Bose97},
closely related to the standard quantum
limit for continuous position detection~\cite{Tittonen99}, and of particular relevance to
experiments aimed at observing superposition states of the mirror~\cite{Penrose03}.
A high mechanical $Q$ is desirable to minimize thermal noise in position detection~\cite{Tittonen99}.
In addition, optical cooling can reduce the mean number of mechanical quanta from an initial value of $\langle n \rangle=k_BT/(h\nu_m)$ by at most a factor of the quality factor $Q$ of the 
resonator~\cite{Wilson08}. Finally, high mirror reflectivity is essential to achieve the high cavity finesse 
required for sensitive position detection, high circulating optical power, and efficient optical cooling~\cite{Kippenberg07}.

The majority of the mirrors used in micromechanical Fabry-Perot cavities~\cite{kleckner10} have incorporated a traditional quarter-wave stack reflector, in which alternating layers of dielectric materials with different refractive indices are deposited. 
The mirrors typically employ 16 - 40 layers, 
and have a thickness in the range of $3.6\,\mu m$ to $9\,\mu m$.  
Mechanical resonators exhibiting room-temperature quality factors in the range of $Q=1000$ to $Q=9000$ have been fabricated directly out of the dielectric stack itself~\cite{Bohm06,Cole08,Groblacher08}; a discussion of the tradeoffs involved with specific dielectric materials is given in~\cite{Cole08}. 
Somewhat higher values of $Q$, up to $Q=15\,000$, have been realized with a 
silicon substrate coated with such a stack~\cite{ArcizetPRL06}.  
Substantially higher values of $Q$, up to $Q=250\,000$ at room temperature, have been obtained by using a silicon mechanical resonator to support a smaller reflecting element~\cite{Tittonen99,kleckner06}. An attractive alternative paradigm is to work with a rigid optical cavity containing a weakly-reflecting silicon nitride membrane~\cite{thompson08,Wilson08}.  This approach exploits the extremely high mechanical quality factor ($Q>10^6$ at room temperature) of silicon nitride membranes, in conjunction with macroscopic cavity mirrors made with well-established methods.

In this work, we demonstrate a micromechanical reflector with an unprecedented combination of high reflectivity,
high mechanical $Q$, and low mass per unit area, by taking an altogether different approach.  
In a dielectric slab patterned as a diffraction grating with a period smaller than the wavelength of the incident light, only zero-order diffraction is allowed.  With appropriate design, the transmitted 
beam can be made arbitrarily small, corresponding to near-unity reflection~\cite{magnusson92,sharon,fan02,
mateus2004,lalanne06,Chang2011}.  Recently, devices employing sub-wavelength grating structures
as reflectors have been demonstrated inside active (VCSEL)~\cite{huang07} and 
passive~\cite{bruckner10} optical resonators.  
Here, we start with a silicon nitride membrane with a reflectivity of $R=27\%$, and by removing material so as to pattern it as a grating, we simultaneously reduce its mass and endow it with a reflectivity exceeding 99.4\%.
We use the patterned membrane as the end mirror in a passive
optical cavity and demonstrate an optical finesse $F>1\,000$ and room-temperature mechanical quality factors as high
as $Q=7.8\times10^5$.  
The mass per unit area of these reflectors is more than an order of magnitude smaller than that of a conventional
mirror with the same reflectivity, and the mechanical quality factor is two orders of magnitude
higher.

\begin{figure}[t] 
  \centering
  \includegraphics[width=3.38in,keepaspectratio]{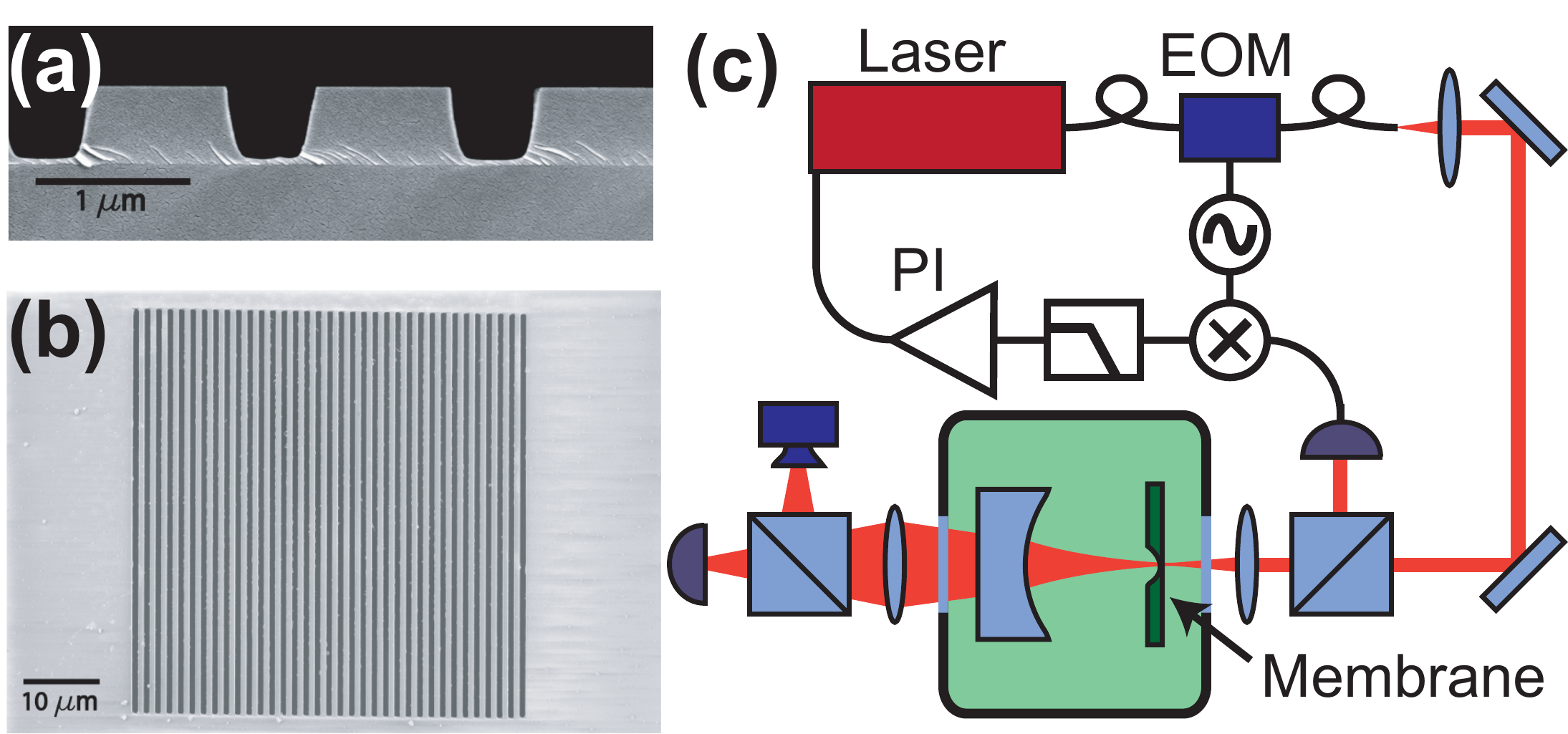}
  \caption{a) Side view of silicon nitride grating structure revealing sloped sidewalls resulting from the etching process.  The substrate is not present in the final devices.  b) One of 81 grating patterns on the membrane, $50\,\mu$m on a side.  c) The membrane containing the grating
structures is used as one mirror of a Fabry-Perot cavity constructed in vacuum.  Cavity transmission
and reflection can both be monitored, and a laser can be locked to a cavity resonance with a PI
(proportional/integral) controller. }
  \label{fig:fig1}
\end{figure}
The design of our silicon nitride grating reflectors started with simulation based on 2D rigorous coupled wave analysis (RCWA)~\cite{Moharam81,MIST}. We chose a relatively long design wavelength of $1.55\,\mu m$ in order to make the device less susceptible to fabrication imperfections.  
Early fabrication attempts revealed sloped sidewalls, as shown in Fig.~\ref{fig:fig1}a.  Simulations showed that the nonrectangular cross section need not
lead to performance degradation, but it must be included in the RCWA calculations.  
The thickness of the silicon nitride film and finger width were chosen in an attempt to make two high-reflectivity modes of the grating coincide~\cite{mateus2004}, in order to make the region of high reflectivity less sensitive to imperfections in fabrication.  

We initiate membrane fabrication by coating silicon wafers with low stress silicon nitride on both sides in a low-pressure chemical vapor deposition system. We then etch out a photolithographically defined patch of silicon nitride from the back surface of the wafer using reactive ion etching (RIE),
and subsequently etch the exposed silicon in KOH to form a silicon nitride membrane.  
We pattern the individual membranes with grating structures by means of electron beam lithography and a subsequent RIE step, in which the electron beam resist is used as a mask.  The membrane used in this work
has dimensions of 1 mm $\times$ 1 mm $\times$ 470 nm, and is patterned with 81 different
grating structures, each 50 $\mu$m on a side; one such structure is shown in Fig.~\ref{fig:fig1}b.
There is a 50 $\mu$m separation between grating patterns, and each grating has a slightly different period and finger width.  All of the results in this paper employ a grating located on a diagonal
of the membrane, 280~$\mu$m from the membrane center, with a period of 1410~nm and a mean finger width of 660~nm.  

In order to characterize the grating reflectors, we constructed a setup allowing an individual grating to be one mirror of a macroscopic Fabry-Perot cavity, as shown in Figure \ref{fig:fig1}c.  
The other cavity mirror is a standard curved dielectric mirror with a radius of curvature $R=25$~mm and reflectivity better than $99.97\%$ over the wavelength range of $1.5\,\mu$m - $1.6\,\mu$m.  
We use a cavity length $L_{cav}$ close to the limit of the stability region given by $L_{cav}<R$~\cite{Kogelnik66}, 
where the waist of the optical mode diminishes rapidly with cavity length. Correspondingly, the cavity free spectral
range $FSR$ is approximately 6~GHz.
The Fabry-Perot cavity is constructed inside a small 
vacuum chamber, with the curved mirror and the grating samples mounted on positioning stages.  Light from a tunable laser is phase-modulated with 
an electro-optic modulator (EOM) in order to derive a Pound-Drever-Hall (PDH) signal, and  
collimated.  
It is then focused through a window in the vacuum chamber and directed onto the grating membrane,
which acts as the input coupler for the Fabry-Perot cavity.   
Light reflected by the cavity is deflected and used to derive a PDH error signal.

\begin{figure}[t] 
  \centering
    \includegraphics[width=3.38in,keepaspectratio]{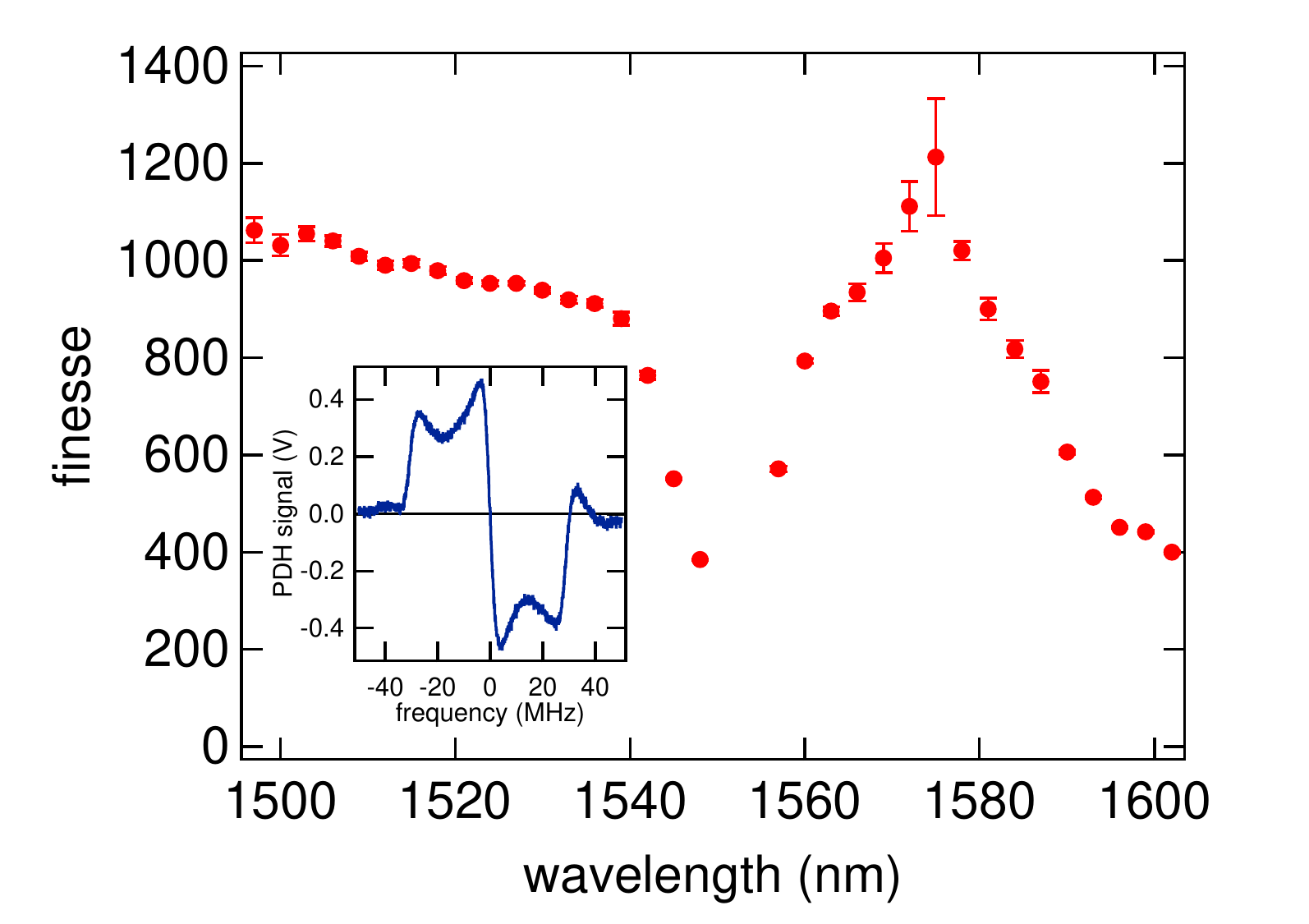}
  \caption{Finesse vs wavelength of cavity in Fig.~\ref{fig:fig1}c.  Inset: PDH signal
  (1 MHz bandwidth) used to lock laser to cavity and transduce mechanical motion.}
  \label{fig:fig2}
\end{figure}
 
The optical response of a cavity employing a particular grating is obtained by sweeping the laser frequency over a cavity resonance while monitoring the cavity transmission and reflection.  
Fig.~\ref{fig:fig2} shows the finesse $F=FSR/\delta \nu_{cav}$ measured over a wavelength range of 1500~nm to 1600~nm, where the cavity FWHM $\delta \nu_{cav}$ is obtained from Lorentzian fits to the transmission.
The inset shows the PDH signal at a wavelength of 1560~nm; the width (peak - valley) of the steep central section is approximately equal to $\delta \nu_{cav}$.
The finesse varies substantially, with maxima at 1500~nm ($F=1060$) and 1575~nm ($F=1200$). 
The finesse of a low-loss cavity is related to the mirror reflectivities $R_i$ by~\cite{Siegman}
\begin{equation}
F=\frac{2\pi}{(1-R_1)+(1-R_2)}.
\end{equation}
where $\Gamma$ describes all other 
round-trip losses, such as absorption, scattering,
and diffraction. A lower bound for the reflection $R_1$ is
thus provided by
\begin{equation}
R_1\ge 1-\frac{2\pi}{F}
\end{equation}
and yields a value of $R_1\ge 99.4\%$ at 1575~nm.  Variations in the reflectivity with wavelength are predicted with RCWA calculations~\cite{mateus2004,MIST}, and we will employ such data in future sample designs. 
Even at a wavelength
of 1548~nm, where the finesse takes its minimum value of $F=385$, the reflectivity is $R_1\ge 98.4\%$.
By comparison, a quarter-wave stack employing conventional dielectrics would need 16 - 34
dielectric layers to achieve a reflectivity of $R= 99.4\%$, would be 7 - 10 times thicker,
and have 16 - 31 times the mass per unit area.  

\begin{figure}[t] 
  \centering
  \includegraphics[width=3.38in,keepaspectratio]{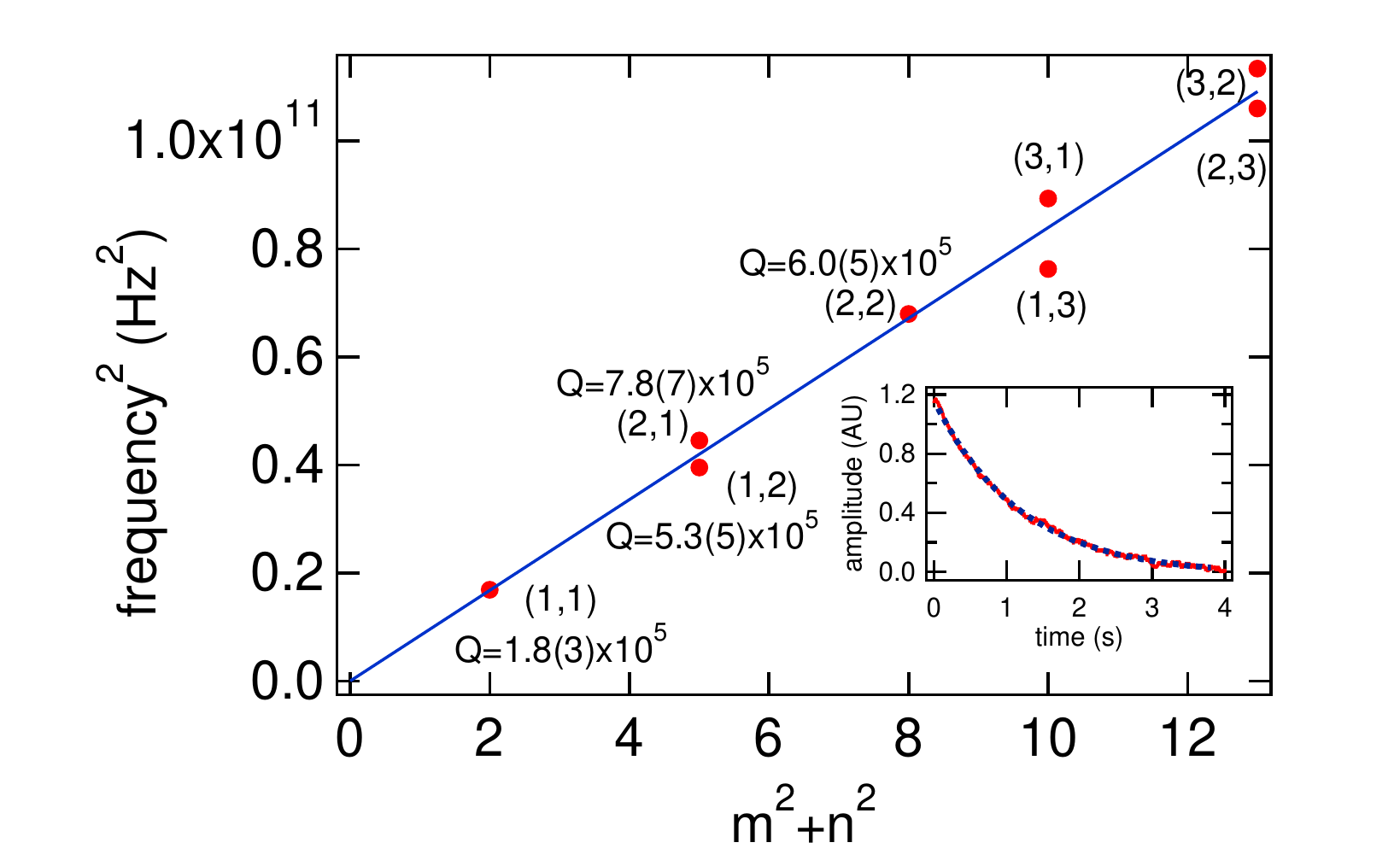}
  \caption{Identification of the eight lowest-frequency mechanical modes, along with mechanical quality 
factors $Q$. The uncertainties given represent the standard deviation
in twenty repeated measurements of $\tau$. Inset: Mechanical ringdown of (2,1) mode at 211~kHz has a time constant of $\tau=1.18$~s, corresponding
to a mechanical quality factor of $Q=7.8\times10^5$. }
  \label{fig:modes}
\end{figure}

We next describe the mechanical characteristics of the grating reflectors.  To this end, we lock a 
laser with a wavelength of 1560~nm to the cavity with a bandwidth of approximately 12~kHz by means of the PDH signal shown in Fig.~\ref{fig:fig2}.  
Analysis of the rapid ($>$12~kHz) variations in the PDH
signal by means of an rf spectrum analyzer reveals hundreds of well-resolved spectral lines in the frequency
range of 130~kHz to several MHz.   
The normal mode frequencies $\nu_{m,n}$ of a uniform square membrane of side $a$, tensile stress $\Sigma$, and density $\rho$ are given by~\cite{Timoshenko}
\begin{equation}
\nu_{m,n}^2=\frac{\Sigma}{4\rho a^2}(m^2+n^2)
\label{eqn: membranemodes}
\end{equation}
where $m$ and $n$ are positive integers.
Fig.~\ref{fig:modes} shows an identification of the eight lowest-frequency modes 
according to eqn.~(\ref{eqn: membranemodes}).  Clearly the degeneracy of modes $(m,n)$ with
the same sum $m^2+n^2$ is broken, a fact readily attributable to the asymmetric stress 
imposed by patterning the membranes with gratings (Fig.~\ref{fig:fig1}b).  
The amplitude of the response is found to decrease substantially at frequencies above 3~MHz;
this is easily understood, as the distance between antinodes~\cite{Timoshenko} of the mode $(m,n)$ is then
smaller than the size of the optical spot on the grating.

An interesting question that arises is whether the motion of individual fingers in the grating
can be observed as well.  
In the present sample, the frequency of the fundamental
finger mode is $\nu_1\approx1.8$~MHz, at which point the density of membrane modes $\nu_{mn}$ is very
high (mean spacing between modes $\approx$~3~kHz).  Thus, even if the motion of a finger were
transduced by the cavity, the response would not be distinguishable from that of a membrane
mode.  The situation is vastly more favorable for
a smaller membrane.  In a square membrane with $a=90\,\mu m$
and finger length $l=50\,\mu m$, for example, the fundamental mode of a finger would fall midway
between the (1,1) and (1,2) modes of the membrane, and thus would be easily identified.  

The rf spectrum analyzer does not have the resolution necessary to measure
the intrinsic linewidths of the individual mechanical modes, so for this purpose we turn to
a time-domain approach.  The membrane is driven to oscillate
at one of its natural frequencies by means of radiation pressure from an intensity-modulated
auxiliary laser, and after extinguishing the excitation, the mechanical ringdown is measured with a resonant PDH probe.  
In this case, the PDH response is demodulated at the mechanical frequency with a lockin
amplifier, and the oscillation amplitude is recorded as a function of time.  A typical 
ringdown measurement, taken on the (2,1) mode with a frequency of 211~kHz, is shown in 
the inset to Fig.~\ref{fig:modes}.
The decay is exponential with a time constant of $\tau=1.18$~s, corresponding to a 
mechanical quality factor of $Q=\pi\nu_{2,1}\tau=780\,000$.  We have measured in this way the mechanical 
quality factors $Q_{m,n}$ of the first four modes, which are given in Fig.~\ref{fig:modes} and
range from $180\,000$ to $780\,000$.   

\begin{figure}[t] 
  \centering
 \includegraphics[width=3.38in,keepaspectratio]{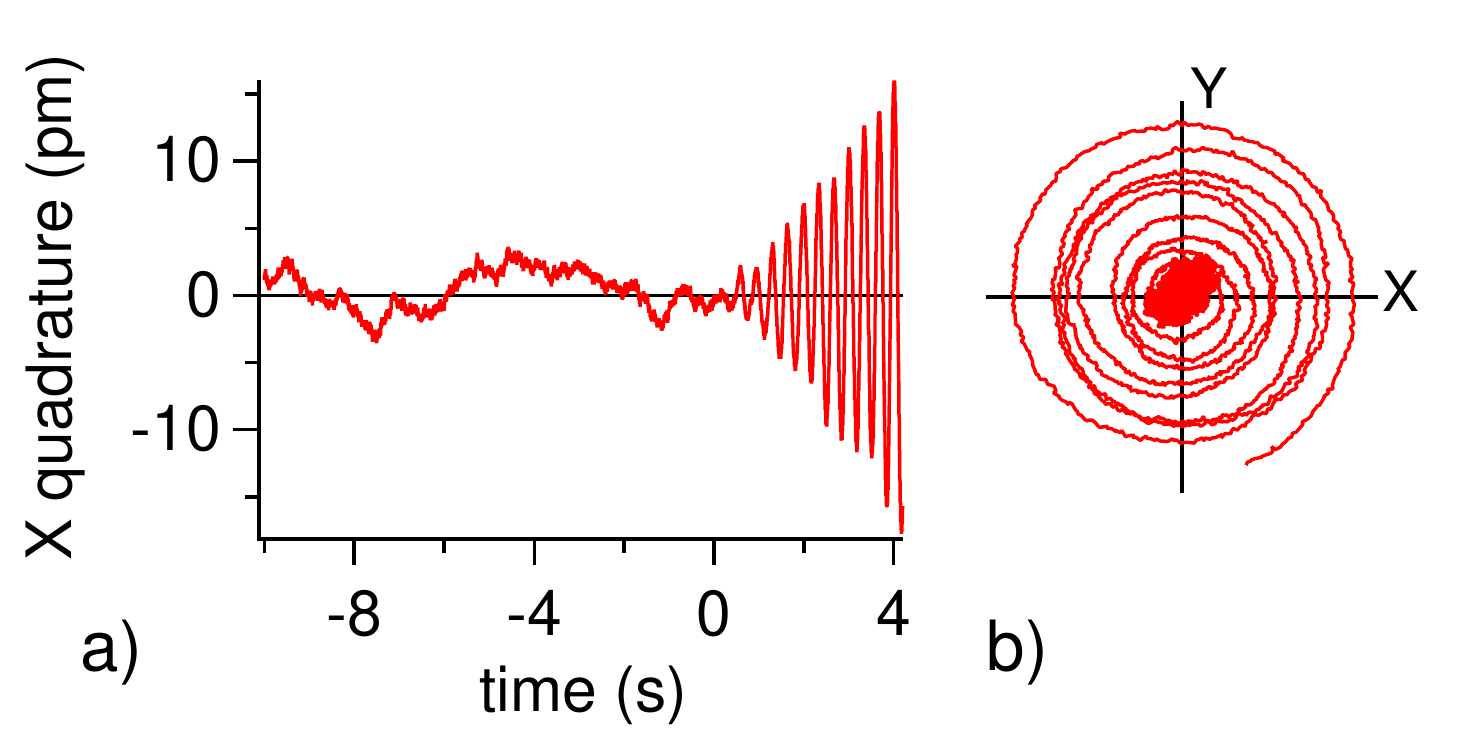}
  \caption{a) Thermal motion is recorded when the laser is resonant with the cavity mode,
changing to oscillatory motion when, at time $t=0$, the laser is detuned to the blue side of the
optical resonance.  b) Polar plot illustrating both quadratures of the demodulated
motion.}
  \label{fig:blueosc}
\end{figure}

Finally, we illustrate qualitatively the case of self-induced oscillation when the 
light injected into the cavity is detuned to the high-frequency (``blue'') side of the 
cavity resonance.  By locking the laser to the cavity with a DC offset in the PDH signal
we force a detuning of $\delta \nu\approx\delta \nu_{cav}/3$ while maintaining an approximately linear response of the signal to the cavity length.
The PDH signal is demodulated at a frequency of 211~kHz 
corresponding to the (2,1) mode.  Fig.~\ref{fig:blueosc}a
shows one quadrature of the demodulated signal where the laser is initially tuned to resonance with the
optical cavity, and then abruptly detuned to the blue at $t=0$.  The
signal for $t<0$ reflects thermal motion, and a detailed analysis confirms
motion with a fluctuating amplitude given by Boltzmann statistics.  In the phasor
representation of both demodulated quadratures shown in Fig.~\ref{fig:blueosc}b, the 
transition between thermal motion with a fluctuating phase and a well-defined oscillatory
behavior is clearly seen, and the fact that the demodulated phase evolves with time (spiral
behavior) is evidence of an optically-induced modification of the oscillation frequency~\cite{Kippenberg07}.

We have demonstrated a micromechanical grating reflector with a reflectivity as high as 99.4\%, a
mechanical quality factor two orders of magnitude higher
than that of mirrors made from conventional quarter-wave stacks, and a mass per unit area more than
an order of magnitude lower.  Fine-tuning the grating parameters and improving the fabrication process for less surface roughness should allow us to build devices with even higher reflectivity.
The mass can be further reduced by using substantially thinner membranes, 
although calculations show that they will be more sensitive to variations in fabrication. 
By working with high-stress silicon nitride we expect to be able to significantly increase the mechanical $Q$ as well~\cite{Wilson08}.  

We are currently studying how the cavity finesse depends upon the optical spot size on
the grating reflector.  This is a matter of interest in its own right~\cite{kleckner10}
and may be crucial to substantially increasing the finesse.  Preliminary finite difference time-domain 
calculations suggest that a finesse of $F=10\,000$
is realistic.  We are also making a number of refinements in our 
experimental apparatus aimed at optical
cooling, and to this end we soon plan to fabricate membranes containing only a single grating structure.  
Such a membrane with sides of $a=90\,\mu$m would have a mass of $m=1.1\times10^{-11}$~kg, 130 times lower than the membrane studied here,
mode frequencies 11 times higher, and, as noted earlier, the response of individual 
fingers within each grating would be easily distinguished from the membrane modes.  
Finally, in addition to their utility in conventional Fabry-Perot cavities, such highly reflective membranes
should be of considerable interest in ``membrane in a cavity'' experiments as well~\cite{thompson08,Wilson08}, which to date have employed membranes with reflectivities below 20\%.

We acknowledge National Science Foundation support through the Physics Frontier Center at the
Joint Quantum Institute.  Research performed in part at the NIST Center for Nanoscale Science and Technology.

%

\end{document}